\documentclass[10pt,aps,prb,twocolumn,preprintnumbers,amsmath,amssymb,floatfix,showpacs,citeautoscript,superscriptaddress]{revtex4-2}
\usepackage[latin1]{inputenc}
\usepackage[T1]{fontenc}
\usepackage[english]{babel}
\usepackage[pdftex]{graphicx}
\usepackage{amsmath}
\usepackage{times}
\usepackage[usenames,dvipsnames,svgnames,table]{xcolor}
\usepackage[colorlinks,plainpages=false,linkcolor=blue,urlcolor=blue,citecolor=blue,pdfpagemode=UseNone]{hyperref}
\usepackage{hyperref}

\renewcommand{\AA}{\text{\r{A}}}

\definecolor{forestgreen(web)}{rgb}{0.13, 0.55, 0.13}

\def\mg{\color{RoyalBlue}}
\definecolor{magenta}{rgb}{0.9,0.0,0.9}
\def\mg{}

\begin{document}

\title
{
\boldmath
Optical properties and electronic correlations in La$_3$Ni$_2$O$_{7}$ bilayer nickelates under high pressure
}

\author{Benjamin Geisler}
\email{benjamin.geisler@ufl.edu}
\affiliation{Department of Physics, University of Florida, Gainesville, Florida 32611, USA}
\affiliation{Department of Materials Science and Engineering, University of Florida, Gainesville, Florida 32611, USA}
\author{Laura Fanfarillo}
\affiliation{Istituto dei Sistemi Complessi (ISC-CNR), Via dei Taurini 19, I-00185 Rome, Italy}
\author{James J. Hamlin}
\affiliation{Department of Physics, University of Florida, Gainesville, Florida 32611, USA}
\author{Gregory R. Stewart}
\affiliation{Department of Physics, University of Florida, Gainesville, Florida 32611, USA}
\author{Richard G. Hennig}
\affiliation{Department of Materials Science and Engineering, University of Florida, Gainesville, Florida 32611, USA}
\affiliation{Quantum Theory Project, University of Florida, Gainesville, Florida 32611, USA}
\author{P.J. Hirschfeld}
\affiliation{Department of Physics, University of Florida, Gainesville, Florida 32611, USA}

\date{\today}

\begin{abstract}
\vspace*{0.2cm}
We explore the optical properties of La$_3$Ni$_2$O$_{7}$ bilayer nickelates %
by using density functional theory including a Coulomb repulsion term.
Convincing agreement with recent experimental ambient-pressure spectra is achieved for $U \sim 3$~eV,
which permits tracing the microscopic origin of the characteristic features.
Simultaneous consistency with angle-resolved photoemission spectroscopy and x-ray diffraction %
suggests the notion of rather moderate electronic correlations in this novel high-$T_c$ superconductor.
Oxygen vacancies form predominantly at the inner apical sites and renormalize the optical spectrum quantitatively,
while the released electrons are largely accommodated by a defect state.
We show that the structural transition occurring under high pressure %
coincides with a significant enhancement of the Drude weight and a reduction of the out-of-plane interband contribution
that act as a fingerprint of the emerging hole pocket.
{\mg We further calculate the optical spectra for various possible magnetic phases including spin-density waves and discuss the results in the context of experiment.}
Finally, we investigate the role of the 2-2 versus 1-3 layer stacking and compare the bilayer nickelate
to La$_4$Ni$_3$O$_{10}$, La$_3$Ni$_2$O$_{6}$, and NdNiO$_2$,
unveiling general trends in the optical spectrum as a function of the formal Ni valence in Ruddlesden-Popper versus reduced Ruddlesden-Popper nickelates.
\vspace*{1.0cm}
\end{abstract}

\maketitle

\section{Introduction}

The recent observation of superconductivity with $T_c \sim 80$~K in pressurized La$_3$Ni$_2$O$_7$ \cite{Sun-327-Nickelate-SC:23, Hou-LNO327-ExpConfirm:23, Zhang-LNO327-ZeroResistance:23}
suggested the bilayer Ruddlesden-Popper compounds as an intriguing new member of the steadily growing family of superconducting nickelates~\cite{Li-Supercond-Inf-NNO-STO:19, Li-Supercond-Dome-Inf-NNO-STO:20, Zeng-Inf-NNO:20, Osada-PrNiO2-SC:20, Osada-LaNiO2-SC:21, Pan-ILSC:22, GoodgeGeisler-NNO-IF:22}
and instantly sparked considerable interest~\cite{Luo-LNO327:23, Gu-LNO327:23, Yang-LNO327:23, Lechermann-LNO327:23, Sakakibara-LNO327:23, Shen-LNO327:23, Christiansson-LNO327:23, Shilenko-LNO327:23, Wu-LNO327:23, Cao-LNO327:23, Chen-LNO327:23, Lu-LNO327-InterlayerAFM:23, ZhangDagotto-LNO327:23, Liao-LNO327:23, Qu-LNO327:23, HuangZhou-LNO327:23, QinYang-LNO327:23, Liu-LNO327-OxVacDestructive:23, ZhangDagotto-RE-LNO327:23, Liu-LNO327-Optics:23, Geisler-LNO327-Structure:23, RhodesWahl-LNO327:23, Wang-LNO327-SC-OxDeficient:23, Yang-LNO327-ARPES:23, Lu-LNO327:23, Wang-LNO327-SC:23, Wang-LNO327-I4mmm:23, Chen-LNO327-SDW:23, ZhengWu-LNO327:23, Kakoi-LNO327:23}. %
Despite these efforts, several aspects remain unclear so far.
Specifically, the pairing mechanism
and its relation to the pressure-driven structural transition
from a $Cmcm$ to an $Fmmm$~\cite{Sun-327-Nickelate-SC:23, ZhangDagotto-LNO327:23, ZhangDagotto-RE-LNO327:23} or $I4/mmm$ space group~\cite{Geisler-LNO327-Structure:23, Wang-LNO327-I4mmm:23},
both involving a suppression of the NiO$_6$ octahedral rotations,
is still intensely debated~\cite{Gu-LNO327:23, Yang-LNO327:23, Lechermann-LNO327:23, Liu-LNO327-OxVacDestructive:23, Lu-LNO327-InterlayerAFM:23, ZhangDagotto-LNO327:23, ZhengWu-LNO327:23}.

The key ingredient to understand superconductivity on a fundamental level is an appropriate electronic structure.
This necessitates a careful assessment of \textit{ab initio} results
and specifically raises a question about the role of electronic correlations in La$_3$Ni$_2$O$_7$.
Angle-resolved photoemission spectroscopy (ARPES) %
reported a two-band Fermi surface at ambient pressure~\cite{Yang-LNO327-ARPES:23}.
Simultaneously, the formation of an additional Ni-$3d_{z^2}$-derived flat band around the zone corner,
$\sim 50$~meV below the Fermi level, has been observed~\cite{Yang-LNO327-ARPES:23}.
Density functional theory analysis of the Drude peak in the in-plane optical conductivity similar to Qazilbash \textit{et al.}~\cite{QazilbashBasov:09} %
suggested that La$_3$Ni$_2$O$_7$ features strong electronic correlations,
placing it in terms of Mott\-ness %
close to the reference superconductor La$_2$CuO$_4$~\cite{Liu-LNO327-Optics:23}.

Intriguingly, optical spectra permit deeper
insight into the electronic structure %
in a more complete energy window %
via the interband transitions.
The availability of recently measured experimental data provides a unique opportunity %
to estimate the correlation effects %
by following this distinct route.

This motivated us to
explore the optical properties of La$_3$Ni$_2$O$_{7}$ bilayer nickelates %
from first principles including a Coulomb repulsion term.
Convincing agreement with the experimental spectrum at ambient pressure is achieved for $U \sim 3$~eV,
which puts us in position to trace the microscopic origin of its characteristic features.
In addition to the in-plane optical conductivity, we also predict the out-of-plane component,
uncovering an unexpectedly strong anisotropy that reverses as a function of frequency.
Simultaneously, Ni~$3d_{z^2}$ energies consistent with recent ARPES results
and accurate lattice parameters are obtained, %
establishing the notion of rather moderate electronic correlations in La$_3$Ni$_2$O$_7$.
Moreover, we provide trends in the optical spectrum due to explicit oxygen vacancies,
which predominantly occur at the inner apical sites.
The released electrons are largely accommodated by an emergent defect state,
in sharp contrast to doping the system.

Subsequently, %
we predict that the structural transition occurring under high pressure is accompanied by 
a significant enhancement of the Drude peak and a reduction of the out-of-plane interband contribution, %
which opens a route to track the proposed changes in Fermi surface topology~\cite{Sun-327-Nickelate-SC:23, Luo-LNO327:23, Gu-LNO327:23, Yang-LNO327:23, Lechermann-LNO327:23, Liu-LNO327-OxVacDestructive:23, ZhangDagotto-LNO327:23}
in future $c$-axis measurements.
{\mg The impact of different magnetic phases including spin-density waves is investigated.}
Finally, we uncover a distinct optical signature of the 2-2 versus 1-3 layer stacking~\cite{Chen-Mitchell-Stacking-LNO327:24, Puphal-Stacking-LNO327:24}
and discuss the bilayer compound in the broader context of related Ruddlesden-Popper and reduced Ruddlesden-Popper nickelates,
identifying fundamental differences between the two families,
but also general trends in the optical response relating to the formal Ni valence.

\begin{figure*}[t]
\begin{center}
\includegraphics[width=\linewidth]{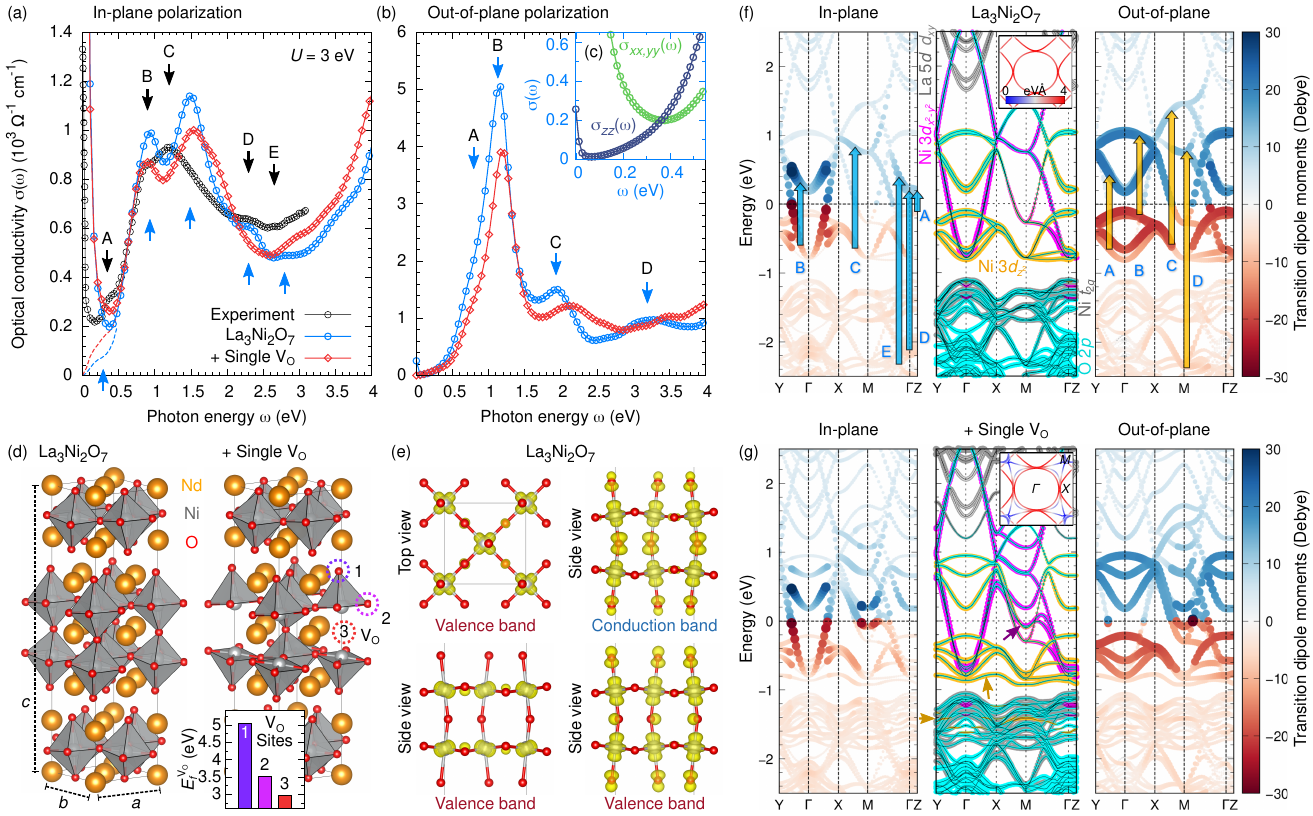}
\caption{\label{fig:OptCond}\textbf{\boldmath Optical properties of La$_3$Ni$_2$O$_7$ at ambient pressure.}
(a)~Optical in-plane conductivity $\sigma_{xx,yy}(\omega)$ and (b)~out-of-plane conductivity $\sigma_{zz}(\omega)$,
comparing DFT$+U$ results for the stoichimetric case (blue) and in the presence of an explicit oxygen vacancy (V$_\text{O}$) (red)
to the experimental data (black~\cite{Liu-LNO327-Optics:23}),
which uncovers very good qualitative and quantitative agreement for the interband transitions.
The dashed lines at low energies represent exclusively the interband contributions.
(c)~Low-energy anisotropy $\sigma_{xx,yy}(\omega)$ versus $\sigma_{zz}(\omega)$ of La$_3$Ni$_2$O$_7$ from panels (a,b).
(d)~Corresponding optimized geometries, %
together with the site-resolved V$_\text{O}$ formation energies. %
(e)~Selected $\Gamma$-point wave functions (absolute square) in La$_3$Ni$_2$O$_{7}$ of the occupied Ni~$3d_{x^2-y^2}$ (top and side view), occupied 'bonding' Ni~$3d_{z^2}$, and empty 'antibonding' Ni~$3d_{z^2}$ states,
revealing a distinct involvement of the inner apical oxygen ions.
(f,g)~Energy- and $k$-resolved transition dipole moments %
$\vert P_{\alpha}^{2}(n,\vec{k}) \vert^{0.5}$ (initial and final states depicted in red and blue, respectively)
together with the orbital-resolved band structure.
Large arrows mark the characteristic transitions shown in panels (a,b).
Small arrows highlight key differences in the band structure induced by the explicit~V$_\text{O}$. %
{\mg The insets show the corresponding Fermi surfaces, colored by the Fermi velocity.}
}
\end{center}
\end{figure*}

\section{Results}

\subsection{Optical spectrum, dipole transition analysis, and impact of oxygen vacancies}

We begin with a detailed analysis of the in-plane optical conductivity $\sigma_{xx,yy}(\omega)$
of La$_3$Ni$_2$O$_7$ at ambient pressure [Fig.~\ref{fig:OptCond}(a); $Cmcm$ space group]
and close comparison with recently measured data obtained from reflectivity experiments~\cite{Liu-LNO327-Optics:23}. %
At least four characteristic interband peaks can be identified in the experimental spectrum:
A small prepeak at around $\omega \sim 0.35$~eV~(A),
two dominant peaks at $0.9$~eV~(B) and $1.2$~eV~(C), and a smaller peak at $2.3$~eV~(D).
Interestingly, we find that the simulated spectrum %
obtained for $U = 3$~eV
nicely reproduces these features both qualitatively and quantitatively.
The quantitative agreement (i.e., the peak intensities and peak energies) is particularly good for peaks B and D,
whereas the energy of peak C is overestimated by $0.3$~eV.
Furthermore, close inspection uncovers a small fifth peak at $\omega \sim 2.7$~eV~(E) that is reproduced at $\sim 2.8$~eV.
The simulation even predicts a finite spectral weight near peak~A.
Since the experimental in-plane Drude peak is significantly renormalized~\cite{Liu-LNO327-Optics:23}, %
as one can see by comparing to our simulated result,
such low-energy interband features can be clearly resolved.

The convincing agreement of the simulated spectrum for La$_3$Ni$_2$O$_7$ and experimental observations
puts us in a position to trace the microscopic origin of its characteristic features.
The energy- and $k$-resolved TDMs $P_{\alpha}^{2}(n,\vec{k})$
in conjunction with the orbital-resolved band structure [Fig.~\ref{fig:OptCond}(f)]
allow us to identify spectral peaks~B and~C to in-plane polarized transitions between
Ni~$3d_{x^2-y^2}$ and Ni~$3d_{z^2}$ states. 
The low-energy spectral weight (A) stems from excitations in the immediate vicinity of the Fermi level, particular at finite $k_z$,
which are promoted by the substantial octahedral rotations. %
At higher energies, transitions from the O~$2p$ valence states to the Ni~$3d_{z^2}$ states can be observed (D, E).

The predicted out-of-plane optical conductivity $\sigma_{zz}(\omega)$ [Fig.~\ref{fig:OptCond}(b)],
while smaller for low energies as expected in a layered system (see inset),
exceeds the in-plane components substantially for $\omega > 0.35$~eV %
and uncovers a strong anisotropy of the optical response.
The spectrum features a very pronounced peak at $\sim 1.1$~eV that consists of at least two distinct contributions~(A, B),
a smaller peak at $\sim 1.9$~eV~(C),
and a broad peak around $3.2$~eV~(D).
The Drude peak is rather negligible.
We attribute these features to strong transitions between
the occupied 'bonding' Ni~$3d_{z^2}$ states and empty 'antibonding' Ni~$3d_{z^2}$ states~(A, B)
as well as unoccupied Ni~$3d_{z^2}$--Ni~$3d_{x^2-y^2}$--O~$2p$ hybrid states around the $M$~point~(C),
but also to excitations from the O~$2p$ valence states to Ni~$3d_{x^2-y^2}$ states~(D) [Fig.~\ref{fig:OptCond}(f)].
Interestingly, comparison of the TDMs thus 
reveals that light with varying polarization excites transitions between highly distinct electronic states
in these novel nickelate compounds.

Figure~\ref{fig:OptCond}(e) visualizes the $\Gamma$-point wave functions of selected states in La$_3$Ni$_2$O$_7$
that contribute prominently to the optical spectrum.
We see that the states extend over the entire bilayer due to the substantial Ni~$3d$-O~$2p$ hybridization.
In the Ni-$3d_{x^2-y^2}$-derived states in the valence band, only the basal oxygen ions are involved.
In stark contrast,
the Ni-$3d_{z^2}$-derived states in the valence and conduction band show a strong hybridization with the apical oxygen ions,
but simultaneously minor contributions from the basal oxygen ions.
Notably, the charge density at the inner apical oxygen site is considerably higher for the  final than for the  initial states;
therefore, even an excitation with out-of-plane polarization
is associated with a certain degree of charge transfer. %
This is consistent with the band structure [Fig.~\ref{fig:OptCond}(f)],
which uncovers an increasing involvement of oxygen in the Ni~$e_g$ states with increasing energy.

A further interesting unknown at the present time
is the role of oxygen vacancies (V$_\text{O}$'s)
in the physics of bilayer nickelates~\cite{Liu-LNO327-OxVacDestructive:23, Geisler-LNO327-Structure:23, Chen-LNO327-SDW:23}.
It has been reported that the normal-state properties of La$_3$Ni$_2$O$_7$ depend sensitively on the oxygen content;
in particular, oxygen appears to control the metallicity of the samples~\cite{Wu-LNO327:01,Taniguchi-LNO327:95,LNO-327-Diffraction-Zhang:94}.
In order to obtain a first-principles impression of the trends in the optical spectrum,
we consider an explicit single V$_\text{O}$ in the unit cell.
Surprisingly, we identify inner apical oxygen [site~3; see Fig.~\ref{fig:OptCond}(d)]
as the lowest formation energy site ($E_f^{\text{V$_\text{O}$}} = 2.99$~eV$/$V$_\text{O}$),
{\mg which has been confirmed independently by high-resolution electron ptychography~\cite{LNO-327-Dong-VO-Visualization:24}.}
Still, this formation energy is higher than e.g.\ in perovskite LaNiO$_3$
($\sim 2.8$~eV \cite{LNO-OxVac-Beigi:15}). %
The formation energies of basal V$_\text{O}$'s (site~2, %
$3.53$~eV$/$V$_\text{O}$)
and outer apical V$_\text{O}$'s (site~1, %
$5.06$~eV$/$V$_\text{O}$)
are even further enhanced
and reach values reminiscent of SrTiO$_3$~\cite{GeislerPentcheva-InfNNO:20, SahinovicGeisler:21},
which suppresses these defects exponentially. %

Figure~\ref{fig:OptCond}(a) shows that the presence of an explicit V$_\text{O}$ leaves the overall structure of the in-plane optical spectrum invariant, but
enhances the spectral weight below $0.6$~eV and reduces the intensity of the two main peaks B and C,
rendering values closer to experiment.
Furthermore, the right slope of peak~C becomes less steep and the spectral weight beyond $\omega > 2.6$~eV is increased,
which also brings the simulated curve closer to experiment.
Simultaneously, peak~D can no longer be resolved,
which indicates that the present V$_\text{O}$ concentration (corresponding formally to La$_3$Ni$_2$O$_{6.75}$)
is higher than in the experimental sample.
For out-of-plane polarization [Fig.~\ref{fig:OptCond}(b)],
the intensity of peak~B is significantly reduced,
while peaks~C and~D are broadened and shifted to higher energies.

Inspection of Fig.~\ref{fig:OptCond}(g) shows that
the overall band structure and the TDMs
stay similar to the stoichiometric case,
particularly in the ideal bilayer without V$_\text{O}$.
Interestingly, two defect states with strong Ni~$3d_{z^2}$ character emerge at $-0.8$ and $-1.5$~eV.
One of these states is split off from the conduction band
and accommodates the two released electrons. %
Slight variations of $E_\text{F}$ can be attributed to states with Ni~$3d_{x^2-y^2}$ character around the $M$~point,
which experience a lifting of the degeneracy %
along the Brillouin zone boundary in the defective bilayer.
{\mg Concomitantly, the Fermi surfaces show a strongly reconstructed shape and reduced Fermi velocity (i.e., increased resistivity; blue colors) in the defective bilayer, particularly near the $M$~point [Fig.~\ref{fig:OptCond}(g)].}

This mechanism confines
the impact of V$_\text{O}$'s largely to the defective bilayers, %
while the remaining system shows an electronic structure close to the stoichiometric compound at only modest electron doping.
This is reflected in the renormalizations of the optical spectrum, which we find to be moderate
in view of the rather high V$_\text{O}$ concentration considered here.
We therefore speculate that
the general nesting properties of the stoichiometric compound are relatively robust
and may carry over to the Fermi surface in slightly oxygen-deficient samples,
resulting in a similar superconducting pairing.

\begin{figure*}
\begin{center}
\includegraphics[width=\linewidth]{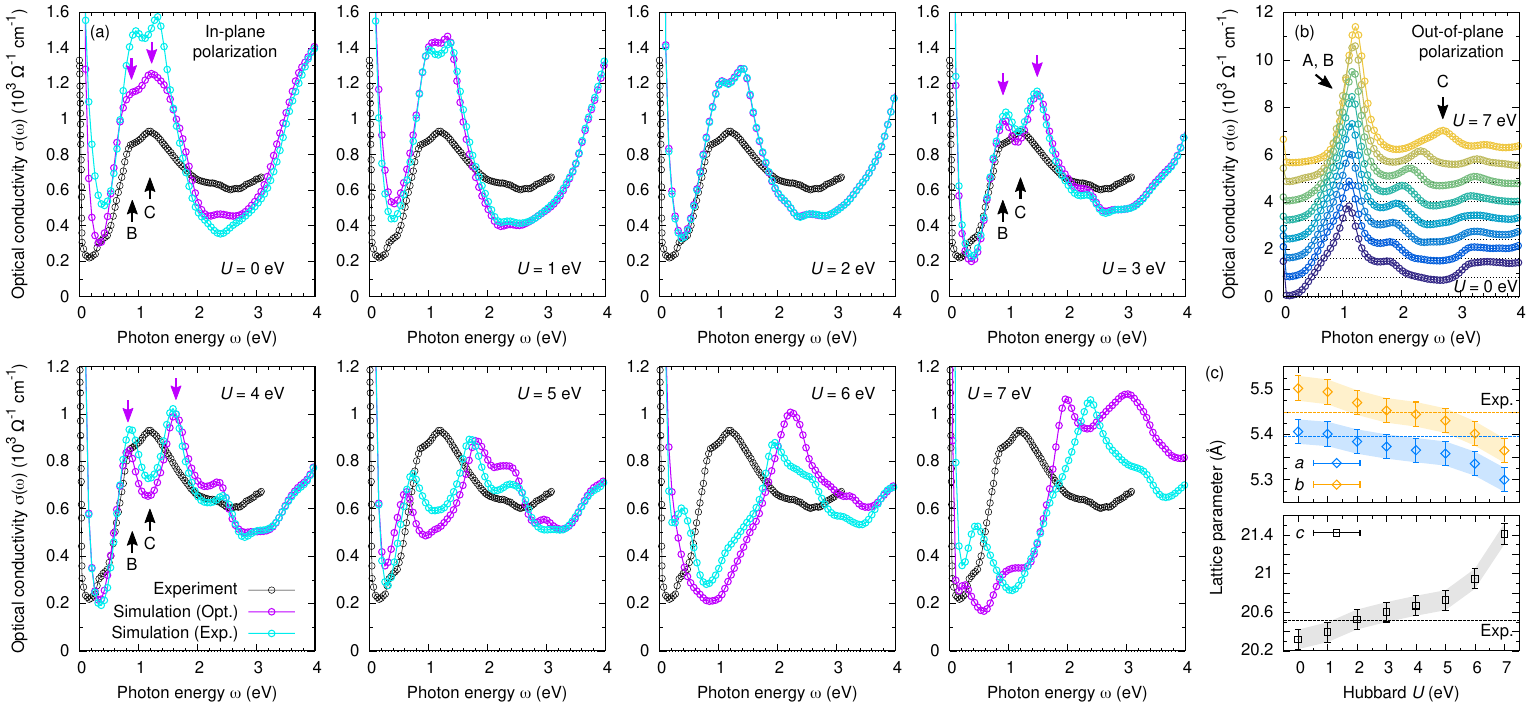}
\caption{\label{fig:UDep}\textbf{\boldmath Correlation dependence of the optical spectrum of La$_3$Ni$_2$O$_7$ at ambient pressure.}
(a)~Optical in-plane conductivity $\sigma_{xx,yy}(\omega)$ %
as a function of the on-site Coulomb repulsion parameter,
exhibiting strong variations with $U$ and very good quantitative agreement with experimental data (black~\cite{Liu-LNO327-Optics:23}) around $U \sim 3$~eV.
In addition to spectra obtained for consistently optimized lattice parameters (magenta),
the panels show results for the experimental cell geometry (cyan). %
(b)~Evolution of the corresponding out-of-plane component $\sigma_{zz}(\omega)$. %
(c)~Optimized lattice parameters ($\pm 0.5\%$ error bar) compared to x-ray diffraction data~\cite{LNO-327-Diffraction-Zhang:94},
demonstrating very good agreement for $U \sim 1.5$-$3.5$~eV. %
}
\end{center}
\end{figure*}

\subsection{Correlation dependence of the optical spectrum} %

Next, we investigate how the ambient-pressure optical spectrum of La$_3$Ni$_2$O$_7$
varies with the Coulomb repulsion parameter,
and estimate the strength of the electronic correlations in this system
by comparison to recent experimental data.

Figure~\ref{fig:UDep}(a) shows that $U$ has a very strong impact on the in-plane optical conductivity, %
affecting both its intensity and shape.
Specifically, we observe
a reduction in overall intensity from $U=0$ to $5$~eV, probably due to decreasing TDMs,
and a simultaneously raising energy of peak~C
due to an increasing separation of the Ni~$3d_{z^2}$ and Ni~$3d_{x^2-y^2}$ states (see Supplementary Information).
For higher $U$ values, the energy of peak~B is concomitantly decreased.

We identify the best overall agreement with recent experimental data~\cite{Liu-LNO327-Optics:23}
for $U=3$~eV [Fig.~\ref{fig:UDep}(a)].
While the B-C energy difference is even more accurately described for yet smaller $U$ values,
the peak intensities are up to twice as high as the experimental curve,
whereas the spectral weight for photon energies between $\omega \sim 2$-$3$~eV is considerably underestimated.
Good quantitative agreement is also found for $U=4$~eV,
albeit with overestimated B-C energy difference.
For $U \geq 5$~eV,
we find a clear disagreement with experiment that particularly manifests in the strongly overestimated B-C energy difference.
This poses an upper limit to the static correlation effects in La$_3$Ni$_2$O$_7$.

Additionally, Fig.~\ref{fig:UDep}(b) displays the correlation dependence of the out-of-plane component $\sigma_{zz}(\omega)$.
The characteristic peak around $\sim 1.1$~eV %
exhibits only a minor energy shift %
and predominantly becomes narrower with increasing $U$.
In sharp contrast, peak~C follows a pronounced nonlinear trajectory towards higher energies with increasing $U$,
reflecting the continuous lowering of the occupied Ni~$3d_{z^2}$ states
and the concomitant raise of the  Ni~$3d_{x^2-y^2}$ states,
particularly the empty ones around the $M$~point (see Supplementary Information).
At present, we are not aware of available experimental data for the out-of-plane component of the optical conductivity.
Future measurements, particularly for peak~C, would provide intriguing additional information about the correlation effects.

The quality of the estimated $U$~value can be tested by analyzing the predicted cell geometry.
Figure~\ref{fig:UDep}(c) shows that
the in-plane lattice parameters~$a$, $b$ decrease monotonically from $5.41$, $5.50~\AA$ ($U=0$~eV) to $5.30$, $5.36~\AA$ ($7$~eV).
In sharp contrast, the out-of-plane lattice parameter~$c$ increases simultaneously from $20.32$ to $21.41~\AA$.
Consistent with the analysis of the optical spectrum,
we find that $U \sim 1.5$-$3.5$~eV renders close agreement (within $\pm 0.5\%$ error bars)
with the experimental lattice parameters $a=5.396$, $b=5.449$, $c=20.516~\AA$ from x-ray diffraction (XRD)~\cite{LNO-327-Diffraction-Zhang:94, Geisler-LNO327-Structure:23}.

Moreover, recent angle-resolved photoemission spectroscopy (ARPES) on La$_3$Ni$_2$O$_7$
identified the formation of Ni-$3d_{z^2}$-derived states $\sim 50$~meV below the Fermi level~\cite{Yang-LNO327-ARPES:23}.
For $U=3$~eV ($4$~eV), we observe the Ni~$3d_{z^2}$ states
\mbox{$\sim 60$~meV} ($150$~meV) %
below the Fermi energy (see Supplementary Information), %
{\mg an important agreement given current debates about the electronic structure of bilayer nickelates.}
This corroborates our conclusion in favor of $U \sim 3$~eV.

By comparing the Drude weight obtained from experiments to DFT calculations,  %
previous work found that La$_3$Ni$_2$O$_7$ features strong electronic correlations
that significantly reduce the kinetic energy of the electrons
and place the bilayer nickelate in terms of Mott\-ness
close to the %
parent compound of cuprate superconductors La$_2$CuO$_4$~\cite{Liu-LNO327-Optics:23}.
The present quantitative analysis of the optical \textit{interband} transitions,
which reflect the relative energies of the active Ni~$e_g$ orbitals
and their distance to the O~$2p$ states in the valence band,
as well as the consistency with ARPES and XRD  %
rather establish the notion of moderate electronic correlations in this novel high-$T_c$ superconductor.

{\mg The rare-earth nickelates~\cite{RENickelateReview:16}
including the infinite-layer compounds~\cite{LeePickett-Inf-LNO:04, Botana-Inf-Nickelates:19, GeislerPentcheva-NNOCCOSTO:21, Geisler-Rashba-NNOSTOKTO:23}
are known to exhibit a more covalent nature and distinct Ni-O hybridization than related cuprates.
Consequently, the latter are often described by higher Hubbard-$U$ values such as $6.5$~eV~\cite{Anisimov:91, ZhongKosterKelly:12}
than nickelate systems ($1$-$4$~eV~\cite{May:10, Chakhalian:11, Gibert-LNO-LMO:12, Liu-NNO:13, KimHan:15, Geisler-LNOSTO:17, GeislerPentcheva-LNOLAO:18}).
It has been reported that the bilayer nickelates show a strong involvement of the oxygen system as well~\cite{ZhangDagotto-LNO327:23, Geisler-LNO327-Structure:23}.
The resulting more delocalized wave functions rationalize the trend of reduced electronic correlations relative to the paradigmatic cuprate superconductors.}

\subsection{Evolution of the optical spectrum under high pressure}

\begin{figure}
\begin{center}
\includegraphics[width=\linewidth]{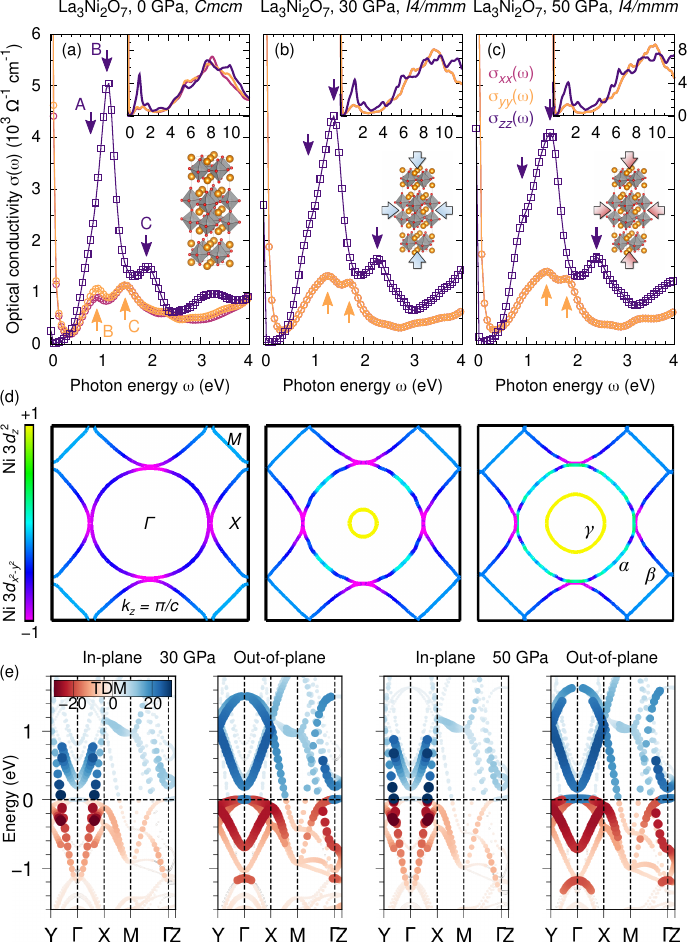}
\caption{\label{fig:DFs}\textbf{\boldmath Evolution of the optical spectrum of La$_3$Ni$_2$O$_7$ under high pressure.}
Optical conductivities $\sigma_{xx}(\omega)$, $\sigma_{yy}(\omega)$, and $\sigma_{zz}(\omega)$ of La$_3$Ni$_2$O$_7$
(a)~at ambient pressure,
(b)~at 30~GPa, %
and (c)~at 50~GPa.
The insets show a larger energy range.
{\mg (d)~Corresponding Fermi surfaces, colored by the Ni orbital character $(d_{z^2}-d_{x^2-y^2})/(d_{z^2}+d_{x^2-y^2})$.
(e)~Energy- and $k$-resolved transition dipole moments at 30~and 50~GPa (cf.~Fig.~\ref{fig:OptCond}) provide a microscopic explanation for the changes observed in the optical spectrum.}
}
\end{center}
\end{figure}

Based on these insights,
we now predict the pressure dependence of the optical conductivity of La$_3$Ni$_2$O$_7$
consistently at $U=3$~eV
and thereby track the changes in the electronic structure across the structural phase transition.

Figure~\ref{fig:DFs}(a) shows $\sigma_{xx}(\omega)$, $\sigma_{yy}(\omega)$, and $\sigma_{zz}(\omega)$
for the orthorhombic $Cmcm$ geometry at ambient pressure. %
In sharp contrast to the pronounced in-plane versus out-of-plane anisotropy, %
the in-plane anisotropy $\sigma_{xx}(\omega) \neq \sigma_{yy}(\omega)$ is rather small.
The inset shows that the optical spectrum
exhibits a distinct 'hump' of moderate anisotropy between $\sim 4$ and $12$~eV. %
We assign it predominantly to transitions from O~$2p$ (below $-2$~eV) to La~$5d$ (above $1.8$~eV)  [Fig.~\ref{fig:OptCond}(f)].
Since the corresponding TDMs are relatively low,
its considerable magnitude is a cumulative effect related to a high density of states. %

At 30~GPa [Fig.~\ref{fig:DFs}(b)],
the in-plane anisotropy vanishes due to the orthorhombic-to-tetragonal phase transition.
The in-plane peaks B ($1.3$), C ($1.7$), and D ($2.45$~eV) as well as the out-of-plane peaks A ($0.9$), B ($1.4$), and C ($2.3$~eV) are shifted to higher photon energies.
This can be understood from a pressure-enhanced
energy difference between the lower and upper Ni~$3d_{z^2}$ states
that is accompanied by an increased energy of the Ni~$3d_{x^2-y^2}$ states,
specifically around the $M$~point~\cite{Geisler-LNO327-Structure:23}.
The in-plane peak~A vanishes for finite pressure (not visible) due to the quenched octahedral rotations.
Simultaneously, the Drude peak is considerably intensified
due to the metallization of the Ni~$3d_{z^2}$ states [$\gamma$~pocket, Fig.~\ref{fig:DFs}(d)]:
The DFT$+U$ plasma frequencies increase from $\omega_{p, xx,yy}^2 \sim 9.75$ to $16.65$~eV$^2$ and from $\omega_{p, zz}^2 \sim 0.02$ to $0.20$~eV$^2$ (see Supplementary Information).
The hump experiences a substantial broadening and now extends towards higher energies
owing to the pressure-induced increase of the O~$2p$ band width
(lowering of the energy onset of the O~$2p$ states)
in the valence band~\cite{Geisler-LNO327-Structure:23}.

At 50~GPa [Fig.~\ref{fig:DFs}(c)],
the in-plane peaks B ($1.4$), C ($1.8$), and D ($2.7$~eV) as well as the out-of-plane peaks A ($0.95$), B ($1.45$), and C ($2.4$~eV)
are shifted to even higher energies.
Concomitantly, the out-of-plane Drude weight is further increased ($\omega_{p, zz}^2 \sim 0.31$~eV$^2$).

Interestingly, we observe a pressure-induced intensity reduction of the out-of-plane peak~B,
which directly reflects the partial depletion of the 'bonding' Ni~$3d_{z^2}$ states in the valence band [Figs.~\ref{fig:DFs}(e) and~\ref{fig:OptCond}(f)].
{\mg Simultaneously, a new transition from $\sim -1.2$~eV to these depleted states becomes active [Fig.~\ref{fig:DFs}(e)].}
Together with the Drude peak enhancement,
this allows us to experimentally track the proposed emergence of a hole pocket~$\gamma$
\cite{Sun-327-Nickelate-SC:23, Luo-LNO327:23, Gu-LNO327:23, Yang-LNO327:23, Lechermann-LNO327:23, Liu-LNO327-OxVacDestructive:23, ZhangDagotto-LNO327:23}),
which plays a key role in the suggested $s^{\pm}$ superconductivity mechanism~\cite{Gu-LNO327:23, Yang-LNO327:23, Liu-LNO327-OxVacDestructive:23, Lu-LNO327-InterlayerAFM:23, ZhangDagotto-LNO327:23}.
Therefore, the results promote optical spectroscopy as powerful technique to track transitions of the Fermi surface topology in high-pressure experiments.

{\mg

\begin{figure}
\mg
\begin{center}
\includegraphics[width=\linewidth]{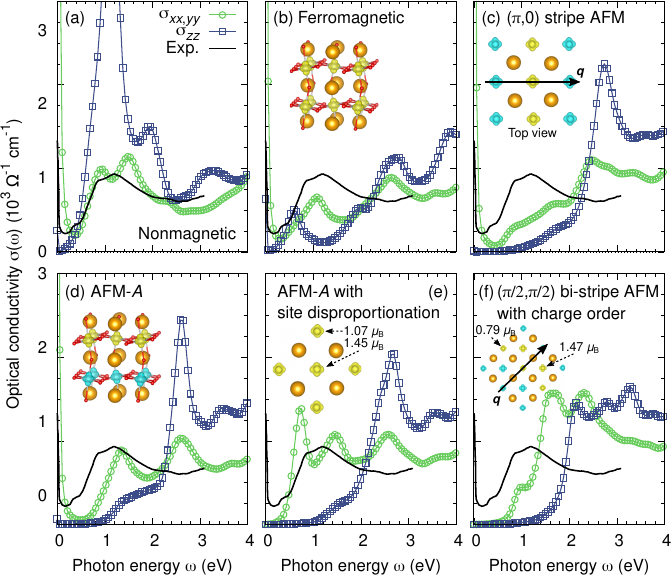}
\caption{\label{fig:Mag}\textbf{\boldmath Optical signature of different magnetic phases in La$_3$Ni$_2$O$_7$ at ambient pressure.}
Compared are $\sigma_{xx,yy}(\omega)$ and $\sigma_{zz}(\omega)$ for
(a)~the nonmagnetic phase,
(b)~ferromagnetic order,
(c)~$(\pi,0)$ antiferromagnetic (AFM) stripe order,
(d)~AFM-$A$ order,
(e)~AFM-$A$ order with in-plane site disproportionation, and
(f)~$(\pi/2,\pi/2)$ AFM charge-spin bi-stripe order as recently suggested SDW model.
The structural models show the spin densities (yellow: positive; blue: negative).
The interlayer coupling is AFM for (c)-(f),
and only a single layer is shown in top view for (c),(e),(f).
The experimental in-plane spectrum obtained at $T \sim 150$~K is displayed as black solid line~\cite{Liu-LNO327-Optics:23}.
}
\end{center}
\end{figure}

\subsection{Optical signature of different magnetic phases}

Recent measurements suggested the presence of a spin-density wave (SDW) in La$_3$Ni$_2$O$_7$ at ambient pressure~\cite{Liu-LNO327-SDW:22, Chen-LNO327-SDW:23}.
However, nuclear diffraction
and nuclear magnetic resonance
studies found no long-range magnetic order~\cite{LNO-327-Diffraction-Ling:00, Fukamachi-LNO327-NMR:01}.
Motivated by this topical discussion,
we explore the optical spectra for different magnetic phases in Fig.~\ref{fig:Mag},
which have been obtained by performing simulations for large supercells.
We find a unique optical signature with strong in-plane versus out-of-plane anisotropy for each magnetic phase.
For instance, the prominent peak in $\sigma_{zz}(\omega)$ (A, B) is shifted to higher energies ($2.6$-$2.8$~eV) for the
$(\pi,0)$ antiferromagnetic (AFM) stripe order suggested in Ref.~\cite{ZhangDagotto-LNO327:23}
and the AFM-$A$ order (with and without site disproportionation,
i.e., an in-plane checkerboard charge and spin modulation~\cite{Geisler-LNO327-Structure:23})
and is less pronounced than in the nonmagnetic case.
For all AFM phases [Fig.~\ref{fig:Mag}(c)-(f)],
the $\sigma_{zz}(\omega)$ Drude peak is negligible or fully quenched,
and the interband onset of $\sigma_{zz}(\omega)$ is found at higher energies than for $\sigma_{xx,yy}(\omega)$.

The $(\pi/2,\pi/2)$ bi-stripe AFM order
superimposed by Ni$^{2+}$/Ni$^{3+}$ charge order [Fig.~\ref{fig:Mag}(f)]
has been proposed recently
by experiment~\cite{LNO-327-Chen-SDW:24}
and theory~\cite{LNO-327-SDW-LaBollita:24}
as good representation of the SDW in La$_3$Ni$_2$O$_7$.
Similar to AFM-$A$ order with site disproportionation,
the $\sigma_{xx,yy}(\omega)$ Drude peak is absent due to the emergence of a finite band gap.

Interestingly,
the experimental curve, which has been measured at $T \sim 150$~K~\cite{Liu-LNO327-Optics:23}
above the SDW transition,
agrees best with the result of our nonmagnetic simulation,
judging from the overall shape and peak positions of the interband spectrum
together with the presence of a Drude peak.
Notably, the sample remains metallic down to low temperatures~\cite{Liu-LNO327-Optics:23}.
These results %
will be helpful in the interpretation of future low-temperature studies.

}

\begin{figure*}
\begin{center}
\includegraphics[width=\linewidth]{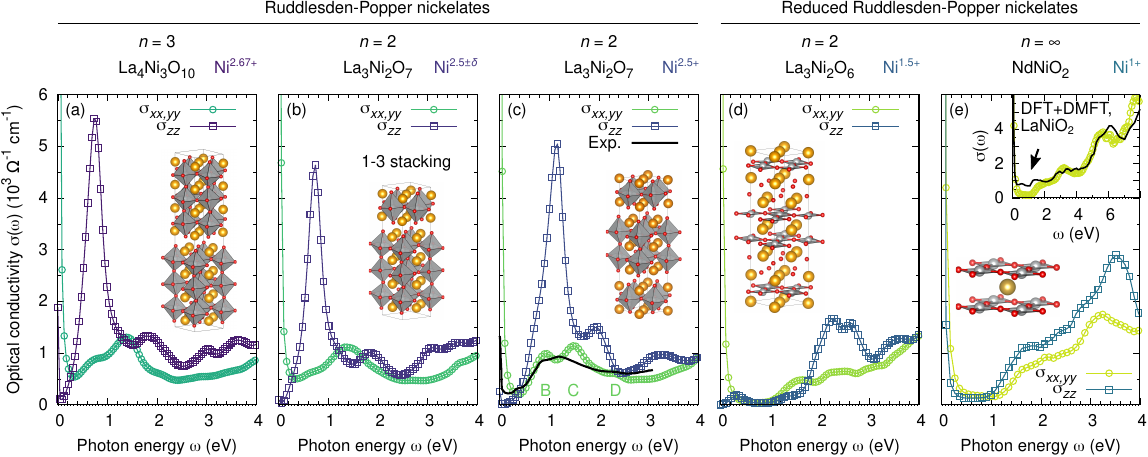}
\caption{\label{fig:RPRRP}\textbf{\boldmath Role of the layer stacking in La$_3$Ni$_2$O$_7$ and comparison with related Ruddlesden-Popper nickelates.}
Optical conductivity (in-plane and out-of-plane) for different members of the Ruddlesden-Popper versus reduced Ruddlesden-Popper nickelates at ambient pressure, ordered by formal Ni valence (indicated by the color palette):
(a)~The $n=3$ trilayer nickelate, (b)~the $n=2$ nickelate with 1-3 stacking, (c)~the $n=2$ bilayer nickelate (experimental spectrum from Ref.~\cite{Liu-LNO327-Optics:23}, black solid line), (d)~the $n=2$ reduced bilayer nickelate, and (e)~the $n=\infty$ infinite-layer nickelate (in-plane DFT$+$DMFT spectrum from Ref.~\cite{Optical-IL-Kotliar:21}, black solid line in the inset).
}
\end{center}
\end{figure*}

\subsection{Role of the layer stacking and comparison with related Ruddlesden-Popper nickelates}

Finally, we discuss the results for the bilayer compound La$_3$Ni$_2$O$_7$ [Fig.~\ref{fig:RPRRP}(c)]
in the broader context of related Ruddlesden-Popper (RP, $A_{n+1}$Ni$_n$O$_{3n+1}$)
and reduced Ruddlesden-Popper (RRP, $A_{n+1}$Ni$_n$O$_{2n+2}$) nickelates.

We observe comparable shapes of the optical conductivity for the three  RP members [Fig.~\ref{fig:RPRRP}(a-c)] shown. %
The trilayer nickelate La$_4$Ni$_3$O$_{10}$ [$n=3$, Fig.~\ref{fig:RPRRP}(a)] has recently been proposed to be superconducting~\cite{Zhu-SC-LNO4310:24, Zhang-SC-LNO4310:24}.
The characteristic out-of-plane peak, which is found to universally stem from transitions between Ni-$3d_{z^2}$-derived quantum-well states [cf.~Fig.~\ref{fig:OptCond}(f)],
is lowered to $\omega \sim 0.75$~eV due to the reduced quantum confinement relative to the bilayer compound.

An interesting variant of the $n=2$ bilayer geometry is the recently observed 1-3 stacking~\cite{Chen-Mitchell-Stacking-LNO327:24, Puphal-Stacking-LNO327:24} [Fig.~\ref{fig:RPRRP}(b)].
We find it to be metastable and $82$~meV$/$f.u. above the 2-2 ground state.
The simulated spectra %
strongly resemble their \mbox{$n=3$} analogs.
In particular, the characteristic out-of-plane peak appears at the same energy as in La$_4$Ni$_3$O$_{10}$, yet with reduced intensity.
Unique manifestations of the single-layer ${\sim 2+}$ Ni states in the spectrum are not observed.

We find the best agreement of the experimental spectrum with our theoretical results for La$_3$Ni$_2$O$_7$ in the bilayer \mbox{2-2} stacking [Fig.~\ref{fig:RPRRP}(c)].
In contrast, in the predicted spectrum for the \mbox{1-3} stacking, the in-plane peak~B is significantly lower than the experimental data and peak~D is not visible at all, whereas peak~C and the spectral weight below $0.6$~eV are overestimated.
We speculate that 2-2 and 1-3 phases may coexist in experimental samples.
Intriguingly, the results establish $\sigma_{zz}(\omega)$ as ideal observable to quantify their relative volume fraction
due to the highly characteristic peak energies.

Complete reduction of the apical oxygen in La$_3$Ni$_2$O$_7$ %
results in the RRP structure La$_3$Ni$_2$O$_6$ \cite{PardoPickett-LNO326:11, ZhangDagotto-LNO326:24} [Fig.~\ref{fig:RPRRP}(d)].
The formal Ni valence jumps from $2.5+$ to $1.5+$ due to the released electrons, %
which we find to be largely accommodated by the Ni~$3d_{z^2}$ states. %
Hence, the spectral weight up to $\omega \sim 2$~eV is strongly suppressed;
specifically, the characteristic peaks in $\sigma_{xx,yy}(\omega)$ and $\sigma_{zz}(\omega)$ vanish.
A double peak appears in $\sigma_{zz}(\omega)$ between $2$-$3$~eV,
which is far less intense than in the RP nickelates and of different origin;
we attribute it to transitions between Ni~$3d_{z^2}$ and La~$5d$ states. %

The $n=\infty$ end member of the RRP series is the infinite-layer geometry [Fig.~\ref{fig:RPRRP}(e)],
in which the Ni~$3d_{z^2}$ states are fully occupied~\cite{Botana-Inf-Nickelates:19, SahinovicGeislerPentcheva:23}.
Here we consider NdNiO$_2$ [Fig.~\ref{fig:RPRRP}(e)],
which hosts superconductivity in film geometry~\cite{Li-Supercond-Inf-NNO-STO:19}.
The in-plane optical conductivity agrees reasonably with earlier %
DFT$+$DMFT simulations for LaNiO$_2$~\cite{Optical-IL-Kotliar:21} (inset),
particular in the interband regime,
which corroborates the validity of our approach.
In addition, we predict the out-of-plane component, which uncovers a further reduced anisotropy.
With respect to La$_3$Ni$_2$O$_6$, the interband onset of $\sigma_{zz}(\omega)$ shifts to lower energies and now coincides with $\sigma_{xx,yy}(\omega)$ at $\sim 1$~eV.
While the characteristic peaks are absent,
we find the higher-energy spectral weight to be enhanced in the RRP versus RP compounds.

These observations mirror fundamental differences in the electronic structure between the two Ruddlesden-Popper families.
The successive occupation of the Ni~$3d_{z^2}$ states
provides each material with its unique optical signature.
This manifests predominantly in the out-of-plane component,
but also for in-plane polarization due to the pronounced Ni~$3d_{x^2-y^2}$-$3d_{z^2}$ coupling (cf.~Fig.~\ref{fig:OptCond}).
The optical spectra of the RP nickelates exhibit characteristic peaks with substantial anisotropy.
In sharp contrast, since the Ni~$3d_{z^2}$ states are largely filled in the RRP compounds,
the corresponding spectra are dominated by the hump,
which is complemented by excitations from Ni~$3d$ to rare-earth~$5d$ states  %
and extends to lower energies due to the higher Fermi level.

\section{Discussion}

The optical properties of La$_3$Ni$_2$O$_{7}$ bilayer nickelates %
were investigated by performing density functional theory simulations including a Coulomb repulsion term.
We analyzed specifically the microscopic origin of the interband transitions,
which reflect the relative energies of the active Ni~$e_g$ orbitals
and their distance to the O~$2p$ states in the valence band,
and uncovered a surprisingly strong anisotropy of the optical response that reverses with frequency.
The optical spectrum was found to be considerably impacted by on-site correlation effects,
and very good agreement with experiment at ambient pressure is attained for $U \sim 3$~eV.
This value simultaneously provides Ni~$3d_{z^2}$ energies consistent with recent angle-resolved photoemission spectroscopy %
as well as lattice parameters that are in close agreement with x-ray diffraction. %

The renormalizations of the optical spectrum due to oxygen vacancies,
which were found to predominantly occur at the inner apical sites,
further improve the agreement with experiment.
Their explicit treatment revealed that
the released electrons are largely accommodated by an emergent defect state,
as opposed to charge doping the sample.

We found that the structural transition occurring under high pressure is accompanied by 
a significant enhancement of the Drude weight
and a concomitant reduction of the out-of-plane interband component,
which act as a fingerprint of the proposed Ni~$3d_{z^2}$ hole pocket formation.
This promotes optical spectroscopy as powerful technique to track %
transitions of the Fermi surface topology in high-pressure experiments.

{\mg
Moreover, we identified the unique optical signature of different magnetic phases in La$_3$Ni$_2$O$_{7}$ at ambient pressure,
including very recent models of spin-density waves.
These insights will be helpful in the interpretation of future low-temperature studies.
}

Furthermore, we discussed the topical question of the \mbox{2-2} versus \mbox{1-3} layer stacking in La$_3$Ni$_2$O$_{7}$.
The out-of-plane optical conductivity emerges as ideal observable to quantify the relative volume fraction
of these possibly coexisting phases due to their highly characteristic peaks.

Finally, we broadened the perspective by comparing the bilayer nickelate
to La$_4$Ni$_3$O$_{10}$, La$_3$Ni$_2$O$_{6}$, and NdNiO$_2$,
which unveiled that each material exhibits its unique optical signature.
We identified fundamental differences between Ruddlesden-Popper and reduced Ruddlesden-Popper nickelates,
but also general trends that relate to the formal Ni valence.

In conclusion, the present comprehensive study
establishes the notion of rather moderate electronic correlations in  La$_3$Ni$_2$O$_{7}$.
The uncovered trends and their microscopic analysis may guide future experiments,
support the quest towards understanding the origin of superconductivity in this high-$T_c$ nickelate, %
and accelerate the discovery of related compounds with enhanced properties.

\section*{Methods}
\subsection*{Density functional theory calculations}

We performed first-principles simulations in the framework of (spin-polarized) density functional theory (DFT~\cite{KoSh65})
as implemented in the \textit{Vienna Ab initio Simulation Package} (VASP)~\cite{USPP-PAW:99, PAW:94}, 
employing a wave-function cutoff of $520$~eV.
Exchange and correlations were described by using the generalized gradient approximation as parameterized by Perdew, Burke, and Ernzerhof~\cite{PeBu96}.
Static correlation effects were considered within the DFT$+U$ formalism~\cite{LiechtensteinAnisimov:95, Dudarev:98},
using $U=3$~eV at the Ni sites unless stated otherwise.

To account for octahedral tilts and oxygen vacancies,
the La$_3$Ni$_2$O$_{7-\delta}$ bilayer nickelates ($\delta = 0, 0.25$) were modeled by using
orthorhombic 48-atom unit cells.
The Brillouin zone was sampled employing
\mbox{$12\times12\times4$} Monkhorst-Pack~\cite{MoPa76} and
$\Gamma$-centered \mbox{$19\times19\times5$} $\Vec{k}$-point grids
in conjunction with a Gaussian smearing of $5$~mRy.
The lattice parameters $a$, $b$ and $c$
as well as the internal ionic positions were accurately optimized in DFT$+U$
under zero and finite external pressure,
reducing ionic forces below $1$~mRy/a.u. %
We proceeded analogously for La$_4$Ni$_3$O$_{10}$, La$_3$Ni$_2$O$_{6}$, and NdNiO$_2$.

The oxygen vacancy formation energies are related to the DFT$+U$ total energies via
\begin{equation*}
    E_f^{\text{V$_\text{O}$}} = E(\text{La$_3$Ni$_2$O$_{7-\delta}$}) - E(\text{La$_3$Ni$_2$O$_{7}$}) + \delta \, \mu_\text{O} \ ,
\end{equation*}
where $\mu_\text{O} = \frac{1}{2} E(\text{O$_2$})$ models the oxygen-rich limit.
The well-known overbinding of gas-phase O$_2$ molecules in DFT necessitates a correction of $E(\text{O$_2$})$,
which we performed such as to reproduce the experimental O$_2$ binding energy of $5.16$~eV~\cite{LNO-OxVac-Beigi:15, GeislerPentcheva-LNOLAO-Resonances:19, GeislerPentcheva-LCO:20, SahinovicGeisler:21, SahinovicGeisler:22, Geisler-VO-LNOLAO:22}.

We obtain the imaginary part of the frequency-dependent dielectric function %
$\varepsilon^{}_{\alpha \beta} (\omega) = \varepsilon^{(1)}_{\alpha \beta} (\omega) + i \, \varepsilon^{(2)}_{\alpha \beta} (\omega)$
from the cell-periodic parts of the Kohn-Sham states $|n, \vec{k} \rangle$
and the corresponding eigenenergies $\epsilon_{n, \vec{k}}$ 
by evaluating~\cite{Gajdos-Bechstedt-LOPTICS:06} %
\begin{align*}
\varepsilon^{(2)}_{\alpha \beta} (\omega) = \frac{4\pi^2 e^2}{\Omega} 
\lim_{q \rightarrow 0} \, \frac{1}{q^2} \sum_{c,v,\vec{k}} 2 w_{\vec{k}} \, \delta( \epsilon_{c, \vec{k}} - \epsilon_{v, \vec{k}} - \omega) \\
\times
\langle c, \vec{k}+q\vec{e}_\alpha | v, \vec{k} \rangle 
\langle v, \vec{k} | c, \vec{k}+q\vec{e}_\beta \rangle .
\end{align*}
Here, $\alpha, \beta \in \{x,y,z\}$ label the different spatial directions,
$\Omega$~denotes the unit cell volume,
and $c$ and $v$ enumerate the conduction and valence band states, respectively.
These interband contributions are complemented by the Drude peak: %
\begin{align*}
\varepsilon^{(2)}_{\alpha \beta} (\omega) = \frac{\Gamma}{\omega} \frac{\omega_{p, \alpha \beta}^2}{\omega^2 + \Gamma^2} .
\end{align*}
We predict the plasma frequency squared $\omega_{p, \alpha \beta}^2$
directly from the DFT$+U$ electronic structure (see Supplementary Information).
In contrast, the lifetime~$\Gamma$ is considerably more difficult to obtain. %
We employ $\Gamma = 10$~meV here,
motivated by recent experimental Drude peak analysis for La$_3$Ni$_2$O$_7$~\cite{Liu-LNO327-Optics:23}
and DFT$+$DMFT results for LaNiO$_2$~\cite{Optical-IL-Kotliar:21}.
Subsequently, the real part of the optical conductivity is obtained via
\begin{align*}
\sigma_{\alpha \beta} (\omega) = \varepsilon_0 \, \omega \, \varepsilon^{(2)}_{\alpha \beta} (\omega) , \quad
\sigma_{xx,yy}(\omega) = \frac{\sigma_{xx}(\omega) + \sigma_{yy}(\omega)}{2} .
\end{align*}
To disentangle the individual contributions to the optical spectrum,
we calculate the transition dipole moments (TDMs),
\begin{align*}
\vec{P}(n,\vec{k} \rightarrow n',\vec{k}') &= \langle n',\vec{k}' | e \hat{\vec{r}} \, | n,\vec{k} \rangle \\
&= \frac{i \hbar e}{(\epsilon_{n, \vec{k}} - \epsilon_{n', \vec{k}'}) \, m}
\langle n',\vec{k}' | \hat{\vec{p}} \, | n,\vec{k} \rangle
\end{align*}
which determine how the system interacts with an electromagnetic wave of a given polarization.
The sum of squares $P^2$ is the transition probability between two states. %
To visualize the TDMs in a compact, yet energy- and $k$-resolved form,
we construct the following cumulative quantities %
for each valence and conduction band state
from the squares of the distinct vector components:
\begin{align*}
P_{\alpha}^{2}(v,\vec{k}) &= \sum_{c} P_{\alpha}^{2}(v,\vec{k} \rightarrow c,\vec{k}) , \\
P_{\alpha}^{2}(c,\vec{k}) &= \sum_{v} P_{\alpha}^{2}(v,\vec{k} \rightarrow c,\vec{k}) .
\end{align*}

\section*{Data availability}

The data is available upon reasonable request to the authors.

\begin{acknowledgments}
We thank Prof.~Roser Valent\'i for helpful discussions.
This work was supported by the National Science Foundation, Grant No.~NSF-DMR-2118718.
L.~F.\ acknowledges support by the European Union's Horizon 2020 research and innovation programme through the Marie Sk\l{}odowska-Curie grant SuperCoop (Grant No.~838526).
\end{acknowledgments}

\section*{Author Contributions}

BG, LF, and PJH conceived of the project. JJH, GRS, RGH, and PJH supervised the research. BG performed the theoretical simulations and corresponding analysis. BG, LF, and PJH wrote the paper. All authors discussed the results and revised the paper.

\section*{Competing Interests}

The authors declare no competing interests.

\end{document}